\documentclass[twocolumn,showpacs,showkeys,preprintnumbers,amsmath,amssymb]{revtex4}
\usepackage{graphicx}% Include figure files
\usepackage{dcolumn}% Align table columns on decimal point
\usepackage{bm}% bold math

\begin{document}

%\nofiles

\preprint{ }

\title{Artificial Decoherence and its Suppression \\
in NMR Quantum Computer
\footnote{
Article presented at QIT13 workshop in November 24, 2005.}
}

\author{Yasushi Kondo$^{1}$, Mikio Nakahara$^{1}$, and Shogo Tanimura$^{2}$\\}

\affiliation{%
$^{1}$Department of Physics, Kinki University, 
Higashi-Osaka 577-8502, Japan\\
$^2$Graduate School of Engineering, 
Osaka City University\\
Sumiyoshi-ku, Osaka, 558-8585, Japan\\
}%

\date{\today}% It is always \today, today,
             %  but any date may be explicitly specified

\begin{abstract}

Liquid-state NMR quantum computer has demonstrated
the possibility of quantum computation and supported 
its development.
Using NMR quantum computer techniques, we observed phase 
decoherence under two kinds of artificial noise fields; 
one a noise with a long period, and the other 
with shorter random period. The first one models 
decoherence in a quantum channel 
while the second one models transverse relaxation.
We demonstrated that the bang-bang control suppresses 
decoherence in both cases.
\end{abstract}


\maketitle

%=====================================================================
\section{Introduction}
%=====================================================================@

Quantum computation currently attracts a lot of attention
since it is expected to solve some of computationally
hard problems for a conventional digital computer~\cite{ref:1}. 
Numerous realizations of a quantum computer have been proposed 
to date. Among others, a liquid-state NMR (nuclear magnetic resonance)
quantum computer is regarded as most successful. 
Demonstration of Shor's factorization algorithm~\cite{VSB01}
is one of the most remarkable achievements.

Although the current liquid-state NMR quantum computer 
is suspected not to be a true quantum computer 
because of its poor spin polarization 
at room temperature \cite{pt},
it still works as a test bench of a working  
quantum computer. Following this concept, we have
demonstrated experimentally using 
an NMR quantum computer that some of theoretical 
proposals really work \cite{qaa,warp}. 
In this contribution, we will show that 
a liquid-state NMR quantum computer can model not only 
a quantum computer  but also the composite system of  
a quantum computer  and its  environment.  
Therefore, one can employ it to test 
the effectiveness of proposed decoherence control methods,
such as a bang-bang control \cite{bb,uchiyama}.
Note that the decoherence control methods are usually 
difficult to be tested because of extremely 
short coherence time in the real system. 

\section{Decoherence}
\begin{figure}[b]
\includegraphics[width=5cm]{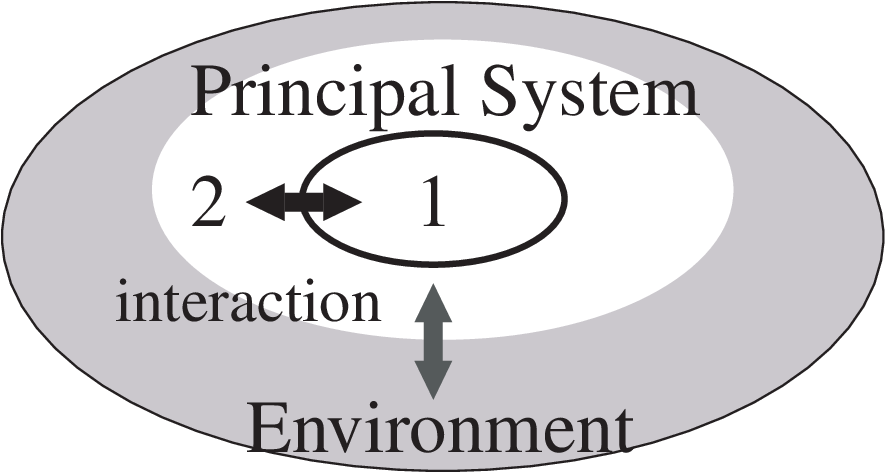}
\caption{
\label{fig1}
System and environment. The system consists of 
subsystems 1 and 2. 
}
\end{figure}

Decoherence is a phenomenon in which a quantum system 
undergoes irreversible change through its interaction 
with the environment. This
is analyzed using the total Hamiltonian %$H_{\rm t}$
\begin{eqnarray}
H_{\rm t} = H_{\rm s} + H_{\rm e}+ H_{\rm se},
\end{eqnarray}
where $H_{\rm s}$ and $H_{\rm e}$, in the absence of $H_{\rm se}$,
determine the system and 
the environment behavior, respectively. On the other hand, 
$H_{\rm se}$ determines the interaction between the system 
and the environment. See, Fig.~\ref{fig1}.
Zurek discussed a simplified model where 
a two-level system (the system) is coupled to 
$n$ two-level systems (the environment) 
through $\sigma_z \otimes \sigma_z$ type interaction \cite{zurek}.

\subsection{Artificial Decoherence}
If the effect of $H_{\rm se}$ is small enough
to be ignored compared with those of $H_{\rm s}$ 
and $H_{\rm e}$ in a certain time scale $\tau$,  
$H_{\rm s}$ can be considered as
\begin{eqnarray}
H_{\rm s} = H_{1} + H_{2} + H_{12} .
\end{eqnarray}
$H_{1}$ and $H_{2}$, if  $H_{12}$ does not exist,
determine the behavior of subsystem 1 and 2, 
respectively, while 
$H_{12}$ determines the interaction between the subsystems.  
See, Fig.~\ref{fig1}.
Therefore, we may regard
the subsystem 1 (2) as the {\it system} 
({\it environment}) in the time scale $\tau$ 
and that the dynamics of the subsystem 1 
can model that of a certain system. 
Zhang {\it et al.} experimentally studied 
the behavior of $^{13}$C-labeled 
trichloroethane, which has three spins,
using NMR techniques \cite{zhang}
and claimed that they studied the decoherence.  

We, however, note that 
a large number of degrees of freedom of 
the subsystem 2 is necessary
to observe a decoherence-like behavior in the
subsystem 1. This condition is not 
satisfied with molecules employed 
in liquid-state NMR quantum computation.
If the degrees of freedom of the subsystem 2 is small,  
a periodic behavior in the subsystem 1 should be
observed instead of an irreversible one.
Teklemariam {\it et al.} introduced a stochastic classical 
field which is acting on the subsystem 2 in order to 
overcome the limitation of the model caused by 
the small degrees of the freedom of the subsystem 2
\cite{Teklemariam}. 
The approach by Teklemariam {\it et al.} can be considered 
as a generation of random noise on the subsystem 1 through 
the subsystem 2 by applying a stochastic classical field 
to the subsystem 2.  

\subsection{Artificial Decoherence in One-Qubit } % System}
Let us consider a molecule containing two spins (qubits)
as a system. The first  qubit is regarded as the subsystem 1 
and the second qubit as the subsystem 2. 
The Hamiltonian, when an individual rotating
frame is assigned to each qubit, is 
\begin{eqnarray}
\label{h_rf}
H_{\rm s} &=& J I_z \otimes I_z,
\end{eqnarray}
where  $I_k = \sigma_k/2$ and $\sigma_k$ is the $k$-th 
Pauli matrix. Note that $H_1=H_2=0$ in this rotating frame. 
We take  a series of $\pi$-pulses acting on
the second qubit as 
a classical field introduced by Teklemariam {\it et al.}
\cite{Teklemariam}.
We create a pseudo-pure state $|00\rangle$ before
starting experiments. Therefore, the Hamiltonian (\ref{h_rf}) is 
equivalent with
\begin{eqnarray}
\label{h_s}
H_{\rm s} &=& J(t) \, I_z. 
\end{eqnarray}
The spin operator $I_z$ appeared in Eq.~(\ref{h_s}) 
denotes the spin of the subsystem 1, while
$|J(t)|=J$ and its sign changes when the $\pi$-pulse 
acts on the spin 2. We assume that the duration of 
a $\pi$-pulses is infinitely short.  
Therefore, we can model 
``a system containing one qubit in a time dependent field'',
where the field strength is constant but its sign changes in
time (telegraphic).

A stochastic classical field is, here, 
a series of $\pi$-pulses acting on
the second qubit randomly in time.

%=====================================================================
\section{Experimental set-up}
%=====================================================================@
A 0.6 ml, 200 mM sample of $^{13}$C-labeled chloroform 
(Cambridge Isotope) in $\mbox{d-6}$ acetone is employed
as a two-qubit molecule and data is taken at room temperature 
with a JEOL ECA-500 NMR spectrometer, whose 
hydrogen Larmor frequency is approximately 500 MHz \cite{jeol}. 
The measured spin-spin coupling constant is $J/2\pi = 215.5$~Hz and 
the transverse relaxation time is $T_2 \sim 7.5$~s 
for the hydrogen nucleus (subsystem 2) and $T_2 \sim 0.30$~s 
for the carbon nucleus (subsystem 1). 
The longitudinal relaxation time is measured to be 
$T_1 \sim 20$~s for both nuclei. 
The duration of a $\pi$-pulses for both nuclei is set to 
$50~\mu$s.  

\subsection{Decoherence in Channel}
Let us consider a flying qubit traveling in a channel, 
as a first example. 
It is assumed that there exists a noise source on a certain position
in the channel, which causes decoherence in the flying qubit. 

\begin{figure}[b]
\includegraphics[width=7cm]{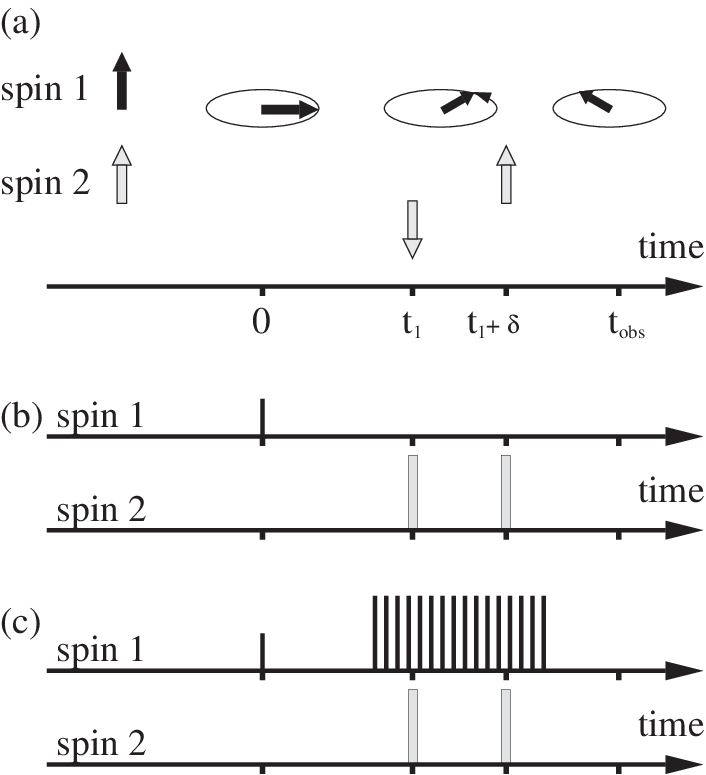}
\caption{
\label{fig2}
Model of ``decoherence in a channel''.  
(a) The dynamics of the spins is schematically shown. 
The spin 1 rotates in the $xy$-plane 
with the angular velocity $-J/2$ except between 
$t=t_1$ and $t=t_1 + \delta$, during which 
the angular velocity is  $J/2$. The spin 2 
is flipped at $t=t_1$ and $t=t_1 + \delta$. The FID signal
of the spin 1 is measured at $t=t_m$. 
(b) The pulse sequences for realizing the spin dynamics 
shown in (a). The short bar indicates a $\pi/2$-pulse
acting on the spin 1, while the long squares are 
$\pi$-pulses acting on the spin 2. 
(c) Pulse sequences compensating the ``decoherence'' 
obtained in (b). The 16 long bars are $\pi$-pulses
acting on the spin 1 with the interval of 0.3~ms.  
}
\end{figure}

\begin{figure}[t]
\includegraphics[width=7.5cm]{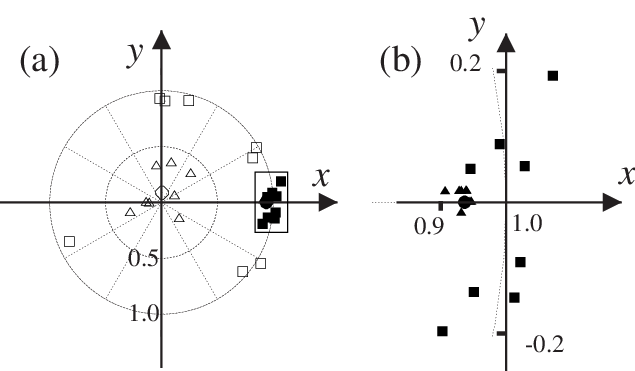}
\caption{
\label{fig3}
Experimental results of ``decoherence in a channel''.
The FID signals measured at $t=t_m$ in Fig.~\ref{fig2}
are shown in the $xy$- (real and imaginary)-plane. 
The solid and open symbols are
with and without the decoherence suppression pulses shown  
in Fig.~\ref{fig2}~(c). 
(a) The open squares denote  
the first 8 FID signals without averaging. 
The 8 open triangles denote the averaged FID signals 
over 16 experiments. The open circle near the
origin denotes the averaged value of all 
($16 \times 8 =128$) the FID signals.
The region marked by the rectangle is enlarged in (b). 
(b) The solid squares denote 
the first 8 FID signals without averaging. The 8  
solid triangles denote the averaged 
FID signals over 16 results. 
The solid square on the $x$-axis shows the averaged
value of all (128) the FID signals.
}
\end{figure}

In order to model the above case, we performed an experiment 
schematically shown in Fig.~\ref{fig2}. 
The pseudopure state $|00\rangle$ 
(or, $|\!\!\uparrow \uparrow \rangle$) is
prepared by the field gradient method \cite{pps}. 
A $\pi/2$-pulse acts on the spin 1 at $t=0$. 
Then, the spin 1 is turned into the $x$-axis in the rotating frame 
and starts rotating in the $xy$-plane
with an angular velocity $-J/2$. 
A pair of $\pi$-pulses are applied to the spin 2 
at $t=t_1$ and $t_1+\delta$ and then the spin 1   
rotates with the angular velocity $J/2$ during this period.  
The FID (free induction decay) signal at $t=t_m$ is measured 
and shown as the open square in Fig.~\ref{fig3}~(a). 
The $x$- ($y$-) component is the real (imaginary) component of 
the FID signal. The signal is normalized so that the data 
point should be on the point $(1,0)$ when $\delta=0$. 
Note that open squares are on a circle of unit radius, 
but with different phases because of different
duration  $\delta$. We randomly 
choose 128 $\delta$s between 0 and $2\pi/J$ and average 
FID signals, as shown in Fig.~\ref{fig3}~(a).
Averaging over all signals gives a smaller 
averaged FID signal than that of each FID signal,
which indicates that decoherence occurred because of the  
noise source in the channel. 

Kitajima, Ban, and Shibata discussed a method to suppress 
the above decoherence \cite{kitajima}. They argued that
a series of $\pi$-pulses acting on the spin 1,
while the spin 1 is under the influence of the noise
source,  should suppresses the above decoherence. 
Their idea is essentially the same as the field inhomogeneity
compensation using the spin echo method \cite{spin_echo}.  
Figure~\ref{fig2}~(c) shows the pulse sequence realizing 
their decoherence suppression proposal. Since we 
do not know the exact position of the noise source in advance,
we apply many (16 here) $\pi$-pulses to the spin 1.
If a noise source exists within this period (equivalently, 
region when the flying qubit is really moving) 
of the 16 $\pi$-pulses, the effect of the noise is 
greatly suppressed. We observed this behavior 
in our experiments, as shown 
in Fig.~\ref{fig3}~(c). The amplitude of each FID signals are
the same and  
the variation of the phases in the $xy$-plane is remarkably
decreased as shown in Fig.~\ref{fig3}~(c). Therefore,
it is clearly seen that decoherence is greatly suppressed.

\subsection{Transverse Relaxation}
Let us model a phenomenon called a transverse relaxation next.
Suppose that there is a spin in a magnetic field $(0,0,B_0)$.
The spin points the $z$-direction in thermal equilibrium.
Then,  let us turn the spin in the $xy$-plane by a $\pi/2$-pulse. 
The spin starts rotating in the $xy$-plane with the 
angular velocity $\omega_0=\gamma B_0$, 
where $\gamma$ is the gyromagnetic ratio of the spin. 
This spin rotation is called a precession and can be 
observed as a FID signal in NMR. 
If there is no relaxation mechanism, it precesses forever
and thus the FID signal does not decay in time. 
The transverse relaxation is a phenomena that the spin 
is still in the $xy$-plane but its FID signal
decreases in time. 
The transverse relaxation is modeled, in the simplest case, 
as a random walk process \cite{rwa,bb} on the circle
shown in Fig.~\ref{fig3}~(a).

\begin{figure}[b]
\includegraphics[width=8cm]{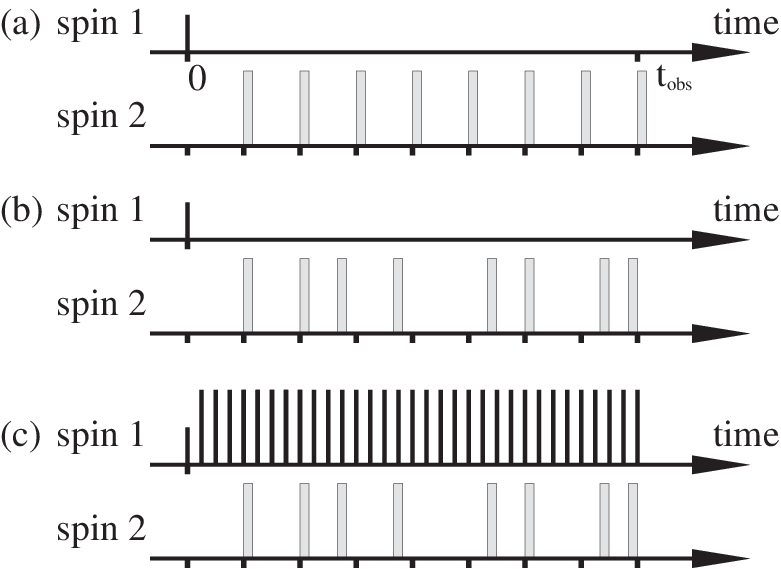}
\caption{
\label{fig4}
Pulse sequences for modeling ``transverse relaxation''.  
The pseudopure state $|00\rangle$ is prepared first. 
The spin 1 is turned into the $x$-axis by the $\pi/2$-pulse 
at $t=0$.  
% The spin 1 points the $x$-direction by a $\pi$-pulse
% at $t=0$.
(a) 
The pulse sequence provides a reference. 
A series of the $\pi$-pulses acts on the spin 2.
The interval of the pulses is fixed to $\Delta =2$~ms.
The sequence shown here yields the solid squares 
in Fig.~\ref{fig5}.  
(b) 
The pulse sequences to realize the artificial
transverse relaxation. The intervals between the pulses
are randomly modulated. See, the text.
% as $\Delta (1+ \alpha \, N_{D})$, 
% where $\alpha$ is a parameter defining the strength 
% of the relaxation and $N_{D}$ denotes a variable 
% which obeys the normal distribution.
The artificial transverse relaxation is obtained
by averaging over various (128) series of 
$\pi$-pulses on the spin 2. 
(c) 
The pulse sequence for realizing the 
bang-bang control. In addition to the pulse 
sequence shown in (b), $\pi$-pulses act on 
the spin 1, of which intervals are $0.5$~ms $ < \Delta$. 
}
\end{figure}

We performed an experiment 
schematically shown in Fig.~\ref{fig4}. 
The pseudopure state $|00\rangle$ is
prepared by the field gradient method \cite{pps}. 
A $\pi/2$-pulse is applied to the spin 1 at $t=0$.
Then, the spin 1 is turned into the $x$-axis in the 
rotating frame and starts rotating with an angular 
velocity $J$. 
The pulse sequence shown in Fig.~\ref{fig4}~(a) 
provides a reference, which is necessary 
because the intrinsic relaxation 
cannot be avoided in the experiments. 
The FID signal is measured at 
$t=t_m$ after a series of $\pi$-pulses acting on the 
spin 2, of which interval is fixed to $\Delta =2$~ms here. 
Note that the number of the $\pi$-pulses on the spin 2 
determines the period while the spin 1 is under 
the influence of time dependent field, 
see Eq.~(\ref{h_s}). The amplitudes of the FID signals 
with various periods are shown as the solid squares. 
When we obtain a faster relaxation than this 
reference, then we can claim that the artificial
relaxation is realized. 

\begin{figure}[t]
\includegraphics[width=7.5cm]{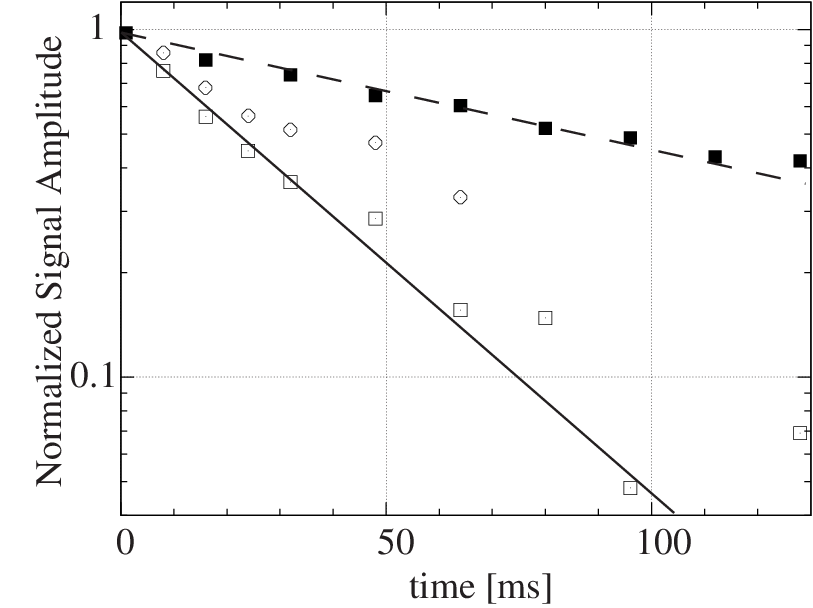}
\caption{
\label{fig5}
Experimental results demonstrating artificial 
``transverse relaxation'' and its suppression.
The solid squares show the amplitude of the 
FID signals of the spin 1 while  
the $\pi$-pulses are acting on the spin 2
every 2~ms.  Those provide a reference for the other
experimental results. 
The open squares show the amplitude of the 
averaged FID signals of the spin 1. 
We averaged 128 FID signals with various random series 
of $\pi$-pulses acting on the spin 2, here. 
The open circles show the amplitude of the 
averaged FID signals as in the case of the open squares 
but with the ``bang-bang'' control. 
The broken (solid) line is the least square fit
of a function $e^{-t/T_2}$ and $T_2 = 128\, (33)$~ms
to the solid (open) squares.
}
\end{figure}

The artificial transverse relaxation is 
realized with the pulse sequence shown 
in Fig.~\ref{fig4}~(b). 
In contrast to (a), the intervals between the pulses
are randomly modulated as $\Delta (1+ \alpha \, N_{D})$, 
where $\alpha$ is a parameter defining the strength 
of the relaxation and $N_{D}$ is a variable which 
obeys the normal distribution. 
We set $\alpha = 0.25 $ in Fig.~\ref{fig5}. 
The artificial transverse relaxation 
is observed when the FID signals with various 
(128 in this experiment) 
series of $\pi$-pulses on the spin 2 are averaged. 
The open squares in Fig.~\ref{fig5} show 
the amplitudes of the averaged FID signals 
with various periods. We  observe 
that the relaxation is faster than the reference case.

The pulse sequence to realize the 
bang-bang control \cite{bb,uchiyama} 
is shown in Fig.~\ref{fig4}~(c). 
In addition to the pulse sequence shown 
in Fig.~\ref{fig4}~(b), regular $\pi$-pulses act on 
the spin 1, of which intervals are 
$0.5$~ms $ < \Delta = 2$~ms. The relaxation with the bang-bang
control is clearly smaller than that without it,
as shown in Fig.~\ref{fig5}. Therefore,
we conclude that the effectiveness of the bang-bang control
is confirmed.  

\begin{figure}[h]
\includegraphics[width=7.5cm]{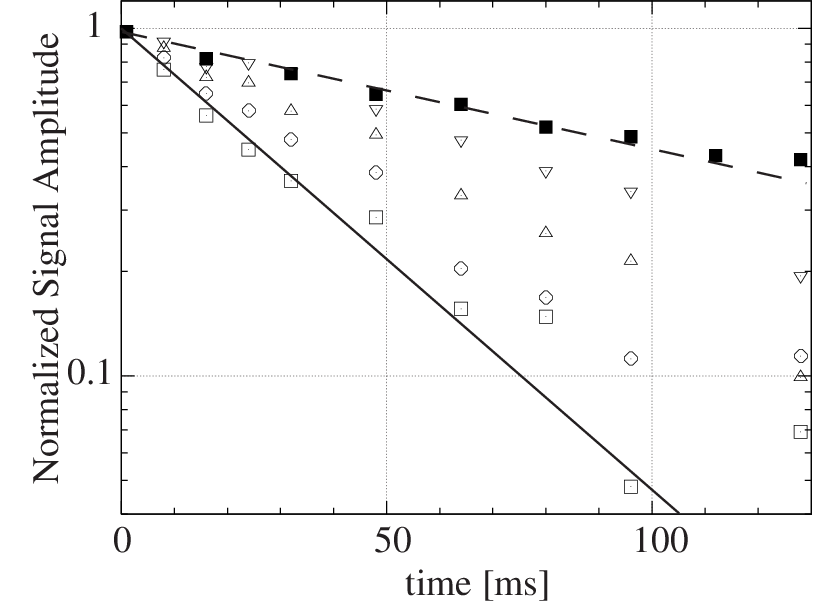}
\caption{
\label{fig6}
Experimental results demonstrating artificial 
``transverse relaxation'' with various $\alpha$ values. 
The open squares, open circles, 
open triangles, and  inverted open triangles show the results  
with  $\alpha =0.25, 0.20, 0.15, 0.10$, 
respectively. The solid squares show the reference
as in Fig.~\ref{fig5}. 
}
\end{figure}

The experimental results, when the parameter $\alpha$ is changed,
is shown in Fig.~\ref{fig6}. 
Note that the strength of the relaxation can 
be controlled by changing the parameter $\alpha$. 
We suspect that there is a non-Markovian
behavior (non-exponential decay of the amplitude of the FID signal) 
when $\alpha=0.10$. We plan to investigate this behavior
further  in near future.  \\ \\

%=====================================================================
\section{Conclusion}
%=====================================================================@
We generated artificial decoherence (relaxations) using 
liquid-state NMR quantum computer techniques. 
The first type of  decoherence takes place in a quantum channel,
while the other is a transverse relaxation. 
Then, the bang-bang control is applied to the qubit
which carries a quantum information and 
we confirmed that it indeed 
suppresses the two types of decoherence.  
Moreover, we have shown that the nature
of the decoherence can be controlled by changing parameters. 

The artificial decoherence thus generated is still simple, but
we can extend our approach further. 
We believe that well controlled artificial decoherence will 
help to understand various types of decoherence in the real world 
and to develop methods to overcome them towards 
physical realization of a working quantum computer. 
 
Extended version of this article with detailed theoretical analysis 
is in progress and  reported elsewhere \cite{cv}.

%=====================================================================
%\section{} %\vspace{-5ex}
%=====================================================================

\end{document}